\documentclass[iop,apj]{emulateapj}
\usepackage{apjfonts}
\usepackage{epsf}
\newcommand{\Msun}{\mathrm{M}_{\odot}}
\newcommand{\Msunyr}{\Msun\ \mathrm{yr}^{-1}}

\def\dim#1{\mbox{\,#1}}

\begin{document}

\slugcomment{Submitted to ApJ}
\shortauthors{Muratov et al.}
\shorttitle{The epoch of Population III stars}

\title{Revisiting The First Galaxies: The epoch of Population III stars} 
 
\author{Alexander L. Muratov \altaffilmark{1}, Oleg Y. Gnedin \altaffilmark{1}, Nickolay Y. Gnedin\altaffilmark{2,3,4}, Marcel Zemp \altaffilmark{1,5}}

\altaffiltext{1}{Department of Astronomy, University of Michigan, 
   Ann Arbor, MI 48109; \mbox{\tt muratov@umich.edu}}
\altaffiltext{2}{Particle Astrophysics Center, Fermi National Accelerator Laboratory, Batavia, IL 60510, USA}
\altaffiltext{3}{Kavli Institute for Cosmological Physics and Enrico Fermi Institute, The University of Chicago, Chicago, IL 60637 USA}
\altaffiltext{4}{Department of Astronomy \& Astrophysics, University of Chicago, Chicago, IL 60637 USA}
\altaffiltext{5}{Kavli Institute for Astronomy and Astrophysics, Peking University, Yi He Yuan Lu 5, Hai Dian Qu, Beijing 100871, China }

\date{\today}

\begin{abstract}

We investigate the transition from primordial Pop III star formation to normal Pop II star formation in the first galaxies using new cosmological hydrodynamic simulations. We find that while the first stars seed their host galaxies with metals, they cannot sustain significant outflows to enrich the intergalactic medium, even assuming a top-heavy initial mass function. This means that Pop III star formation could potentially continue until $z \approx 6$ in different unenriched regions of the universe, before being ultimately shut off by cosmic reionization. Within an individual galaxy, the metal production and stellar feedback from Pop II stars overtake Pop III stars in 20-200 Myr, depending on galaxy mass.

\end{abstract}

\section{Introduction}

The first stars in the universe formed in gas devoid of metals. This exotic environment may have caused the initial mass function (IMF) for Population III stars to be different from the modern day case. Namely, the high Jeans mass of metal-free gas suggests a top-heavy IMF \citep{abel_etal00, bromm_etal99, abel_etal02, yoshida_etal03}. In turn, the feedback processes in the first stars may have been more drastic, prompting the release of extreme amounts of ionizing radiation \citep{tumlinson_shull00, bromm_etal01a} and the occurrence of pair instability supernovae (PISNe) \citep{heger_woosley02}.

To explore the effects of these stars on their host galaxies, we developed a model for Pop III star formation and feedback and implemented it into the adaptive refinement tree (ART) code, as described in a companion paper, (\citealt{muratov_etal13}, hereby Paper I). Pop III stars were modeled to form in gas that was dense, partially molecular, and of primordial composition. Pop III SNe and ionizing radiation feedback were enhanced relative to their Pop II counterparts, and the first PISNe seeded the intrestellar medium (ISM) with metals. We ran a suite of cosmological simulations with this model, and found that the dynamical impact of Pop III feedback depended strongly on the galaxy mass. In agreement with previous work in the field (e.g. \citealt{bromm_etal03, whalen_etal08, wise_etal12a}), we found that PISNe were able to efficiently expel gas and metals from the $M_h$ \textasciitilde $10^6 \, \Msun$ halos expected to host the very first stars \citep{tegmark_etal97}. However, these effects were often temporary, as cosmological inflows of fresh gas restored the baryon fraction to the universal value. The metals, which had previously escaped past the virial radius, also typically fell back into the growing potential wells of the accreting galaxies, leaving the intergalactic medium (IGM) mostly pristine. In galaxies with mass $M_h > 10^7 \, \Msun$, most gas remained bound even after a PISN event, and metals were not ejected past the virial radius. 

Since Pop III stars by definition only form in primordial gas, the large amount of metals released in PISNe leads to the 'self-termination' of Pop III star formation \citep{yoshida_etal04}. According to our findings in Paper I, this self-termination can only be local, as enrichment of the IGM and external halos is rather minimal from single PISNe. Therefore, determining the epoch when Pop III termination becomes universal is a somewhat different question \citep{tornatore_etal07}. Pop III star formation could be relevant for a much longer phase of cosmic history if a Pop III star formed in every pristine halo with $M \gtrsim 10^6-10^8 \, \Msun$ prior to reionization, as the abundance of such halos increases considerably with cosmic time.

Population II star formation can commence in galaxies once they are either sufficiently massive to enable the rapid gas cooling by atomic hydrogen lines, or enriched enough to enable efficient metal cooling \citep{ostriker_gnedin96}. Because the feedback of Pop II stars is weaker than that of Pop III, Pop II star formation should ramp up rapidly in the host galaxy, provided that accretion from filaments continues to bring new supply. However, the relative weakness of the feedback, taken in conjunction with the plethora of primordial sites where Pop III stars may still form, means that the host galaxy, as well as the universe as a whole, are still influenced by Pop III stars for some time after Pop II star formation begins. Though this scenario has already been explored through semi-analytical models(e.g. \citealt{scannapieco_etal03, yoshida_etal04, schneider_etal06}) and numerical simulations (e.g. \citealt{tornatore_etal07, maio_etal10, maio_etal11, greif_etal10, johnson_etal13, wise_etal12a}), understanding this transition quantitatively is relevant for the ability of future observational facilities such as the James Webb Space Telescope (JWST) to observe galaxies dominated by Pop III stars. Studies thus far have shown that the first galaxies generally sit on the brink of detectability by JWST \citep{pawlik_etal11, pawlik_etal12, zackrisson_etal11, zackrisson_etal12}. 

In this paper, we follow the evolution of the galaxies described in Paper I through the epoch of dominance of the first stars. This sample of simulated galaxies spans a range of masses and accretion histories, therefore representing a broad variety of cosmic environments. We study the transition from Pop III to Pop II star formation, and quantify the duration of this epoch. We also explore the effect of cosmic variance, and determine the prevalence and importance of Pop III stars at various cosmic epochs.

\section{Simulations}
A full description of our simulation setup, including the details of both the Pop III and Pop II star formation recipes, is presented in Paper I. Here, we outline the setup only briefly. We perform the simulations with the Eulerian gasdynamics+N-body adaptive refinement tree (ART) code \citep{kravtsov_etal97, kravtsov99, kravtsov03, rudd_etal08, gnedin_kravtsov11}. We use a $256^3$ initial grid with up to 8 additional levels of refinement. For most of our runs, we apply this grid to a 1 $h^{-1}$ Mpc comoving box with the WMAP-7 cosmology ($\Omega_m = 0.28$, $\Omega_\Lambda = 0.72$, $h=0.7$, $\sigma_8 = 0.817$, $\Omega_{b} = 0.046$, $\Omega_{DM}=0.234$). This gives us a dark matter (DM) particle mass $m_{\, DM} = 5.53 \times 10^3 \, \Msun$ and a minimum cell size of 22 comoving pc. We also employ a 0.5 $h^{-1}$ Mpc comoving box, where using the same grid, the DM particle mass is set to $m_{\, DM} = 690 \, \Msun$ and the minimal cell size is 11 comoving pc.

Pop III stars formed in the almost pristine gas with the abundance of heavy elements below the critical metallicity $\log_{10} Z/Z_\odot = -3.5$ \citep{bromm_etal01b}. Cells were allowed to form Pop III stars if the gas density exceeded a threshold $n_{H, min}$, and a molecular hydrogen fraction threshold $f_{H_2, min}$. Through a series of convergence tests, we found that $n_{H, min}=10^4 \dim{cm}^{-3}$ and $f_{H_2, min}=10^{-3}$ were appropriate values for the two thresholds. 

The Pop III prescription was designed to test the maximum possible effect of feedback, relying on an IMF that was top-heavy. Half of the Pop III stars formed as $170 \, \Msun$ particles and were set to explode in PISNe. Each PISN injected $27 \times 10^{51}$ erg of thermal energy and $80 \, \Msun$ of metals into the ISM \citep{heger_woosley02}. The remaining 50\% of Pop III stars formed as $100 \, \Msun$ particles that explode in type II SNe, generating $2 \times 10^{51}$ erg of energy. All Pop III stellar particles emited a factor of 10 more ionizing photons per second than their Pop II counterparts of the same mass \citep{schaerer02, wise_cen09}. A suite of cosmological simulations performed with this model revealed that Pop III stars drastically affected halos with $M_h < 3 \times 10^6 \, \Msun$, but not halos of higher masses. Extended convergence tests revealed that without sufficient mass resolution, it was easy to miss these important dynamical effects. 

In gas that is enriched beyond the critical metallicity, Pop II star formation is modeled according to the molecular-based star formation recipe presented in \citet{gnedin_etal09} and \citet{gnedin_kravtsov11}.

In Table \ref{tab:sims} we list all of the simulations from the suite which we analyze in this paper. Some of the simulations that were used in Paper I for convergence tests and for determining the best parameters for Pop III star formation are not included here. In addition to the three fiducial 1 $h^{-1}$ Mpc boxes (UnderDense$-$\_nH1e4\_fid, UnderDense+\_nH1e4\_fid, and OverDense\_nH1e4\_fid), we study runs with extreme feedback, where we increased PISNe energy (run OverDense\_ExtremeSN) and ionizing photon emission (run OverDense\_ExtremeRad) by a factor of 10. We also include a run with a conservative "low-mass" Pop III IMF that assigns the same feedback parameters to Pop III as for Pop II stellar particles (run OverDense\_LowMass) to account for the present uncertainty in the Pop III IMF (e.g. \citealt{oshea_norman07, greif_etal11}).

 \begin{figure}[t]
\vspace{-0.2cm}
\centerline{\epsfxsize3.5truein \epsffile{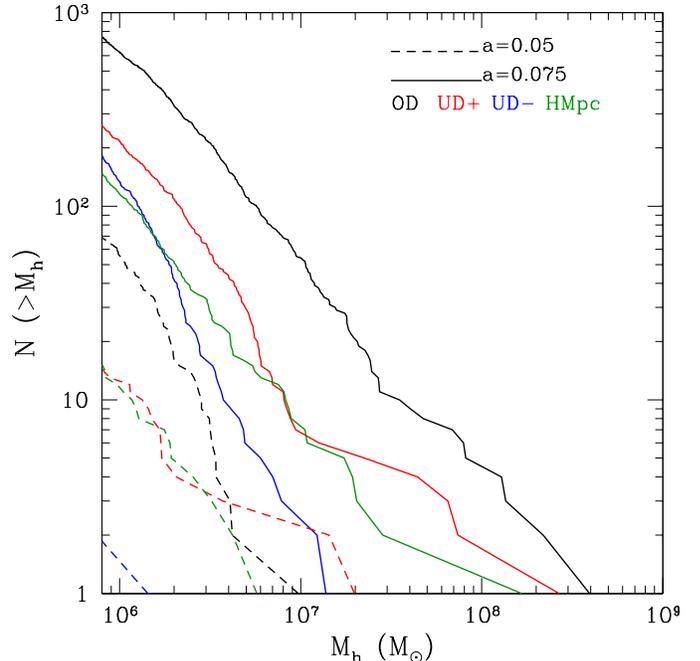}}
\vspace{-0.2cm}
\caption{ Cumulative number of halos vs. mass at two epochs for the three 1 $h^{-1}$ Mpc boxes and one 0.5 $h^{-1}$ Mpc box. The smallest mass plotted corresponds to the earliest halo to form a Pop III star among all of our runs. Box OverDense (black) has an order of magnitude more halos than Box UnderDense+ (red) and Box UnderDense$-$ (blue) across almost the entire range of masses considered. While Box UnderDense+ and Box UnderDense$-$ have similar DC mode values, the former clearly has more massive halos. One Box UnderDense+ galaxy in particular is on par with the most massive galaxies in Box OverDense. The H Mpc box (green, run OverDense\_HMpc\_HiRes) represents only an eighth of the volume of the other boxes, and therefore hosts fewer massive galaxies while still representing an overdense region.}
\vspace{0.3cm}
\label{fig:MF}
\end{figure}

In Paper I, we demonstrated that mass resolution was crucial for capturing Pop III star formation in halos of mass $ M_h$\textasciitilde$10^6 \, \Msun$, and that halos with $M_h < 3 \times 10^6 \, \Msun$ were most susceptible to Pop III feedback. Our fiducial 1 $h^{-1}$ Mpc runs lacked the resolution to study these objects effectively, but using the same grid on a smaller box increased the resolution sufficiently while keeping the computational cost down. For this reason, we also use a 0.5 $h^{-1}$ Mpc box (referred to as H Mpc) for run OverDense\_HMpc\_HiRes, to test the validity of our results in the low-mass regime. Run OverDense\_SL7 employs a super-Lagrangian refinement, allowing for the simulation to more effectively zoom in on overdense regions in primordial galaxies, allowing Pop III stars to form at early times in low mass halos, hence extending the mass range of our sample.

\begin{table*}
\begin{center}
\caption{\sc Simulation runs}
\label{tab:sims}
\begin{tabular}{lrrrrrl}
\tableline\tableline\\
\multicolumn{1}{l}{Run} &
\multicolumn{1}{l}{Base grid} &
\multicolumn{1}{l}{$\ell_{max}$} &
\multicolumn{1}{l}{dx (pc)}  &
\multicolumn{1}{l}{$m_{\, DM} (\Msun)$}  &
\multicolumn{1}{l}{$n_{H, min}($\hspace{-0.05cm}$\dim{cm}^{-3})$}  &
\multicolumn{1}{l}{Description}  
\\[2mm] \tableline\\
\vspace{0.4cm}
\it{Fiducial runs} \\
\vspace{-0.8cm}
\\[2mm] \tableline\\
UnderDense$-$\_nH1e4\_fid & $256^3$ & 8 & 22 & 5500 & 10000 &  Underdense box, no high-mass galaxies, fiducial parameters  \\
UnderDense+\_nH1e4\_fid & $256^3$ & 8 & 22 & 5500 & 10000 & Underdense box, one high-mass galaxy, fiducial parameters \\
\vspace{0.4cm}
OverDense\_nH1e4\_fid & $256^3$ & 8 & 22 & 5500 & 10000 & Overdense box, fiducial parameters \\
\vspace{-0.1cm}
\it{Alternative refinement \& resolution}
\\[2mm] \tableline\\
OverDense\_SL7 & $256^3$ & 8 & 22 & 5500 & 10000 & Super-Lagrangian refinement \\
\vspace{0.4cm}
OverDense\_HMpc\_HiRes & $256^3$ & 8 & 11 & 690 & 10000 & 0.5 $h^{-1}$ Mpc box, high spatial \& mass resolution \\
\vspace{-0.1cm}
\it{Alternative feedback}
\\[2mm] \tableline\\
OverDense\_ExtremeSN & $256^3$ & 8 & 22 & 5500 & 10000 &  Extreme PISNe (Section \ref{sec:extremeFB}) \\
OverDense\_ExtremeRad & $256^3$ & 8 & 22 & 5500 & 10000 &  Extreme Pop III radiation field (Section \ref{sec:extremeFB}) \\
OverDense\_LowMass & $256^3$ & 8 & 22 & 5500 & 10000 & Pop III IMF and feedback mirror Pop II (Section \ref{sec:LowMass}) \\
\\[2mm] \tableline
\end{tabular}
\end{center}
\vspace{0.3cm}
Column 1.) Name of the run; 2.) Base grid, number of DM particles, number of root cells; 3.) Maximum number of additional levels of refinement; 4.) Minimum cell size at the highest level of refinement in comoving pc; 5.) DM particle mass in $\Msun$; 6.) Minimum H number density for Pop III star formation in $\dim{cm}^{-3}$; 7.) Further description of the run.
\vspace{0.3cm}
\end{table*}

\section{Results}
\label{sec:results}

\subsection{ Cosmic Variance }

Using the DC mode formalism for the generation of initial conditions \citep{sirko05, gnedin_etal11} allows us to test several representative regions of the universe without sacrificing resolution, as would be needed were we to simulate a larger cosmological volume. With a single parameter that stays constant over time in a given simulation box, $\Delta_{DC}$, we encode the amplitude of density fluctuations on the fundamental scale of the box. Although many studies have been done to understand the effects of cosmic variance on the dark matter halo mass function \citep{tinker_etal08}, studying it in hydrodynamic simulations is significantly more difficult. Here we present an analysis of the variance of the three different 1 $h^{-1}$ Mpc boxes in our study. For reference, the DC mode values are $\Delta_{DC} = -2.57, -3.35$, and $4.04$ in Box UnderDense$-$, UnderDense+, and OverDense respectively. The H Mpc box has a DC mode of $\Delta_{DC}= 5.04$. Large positive values of $\Delta_{DC}$ indicate an overdense region.

The first Pop III stars form around scale factor $a \approx 0.047$ ($z\approx20.3$) in both Box UnderDense+ and Box OverDense. On the other hand, Box UnderDense$-$ stalled significantly, and no star formation occurs until $a \approx 0.073$ ($z\approx12.7$), translating to a time difference of 165 Myr. This wide range demonstrates immediately that the epoch when Pop III star formation begins is a strong function of local overdensity. If voids like the one represented by Box UnderDense$-$ are indeed not enriched by external sources, Pop III stars could have existed in these voids until the end of cosmic reionization, at epochs that will be probed by JWST \citep{hummel_etal12} and the Large Synoptic Survey Telescope (LSST) (\citealt{trenti_etal09}; but see \citealt{pan_etal12b}).

\begin{figure}[t]
\vspace{-0.2cm}
\centerline{\epsfxsize3.5truein \epsffile{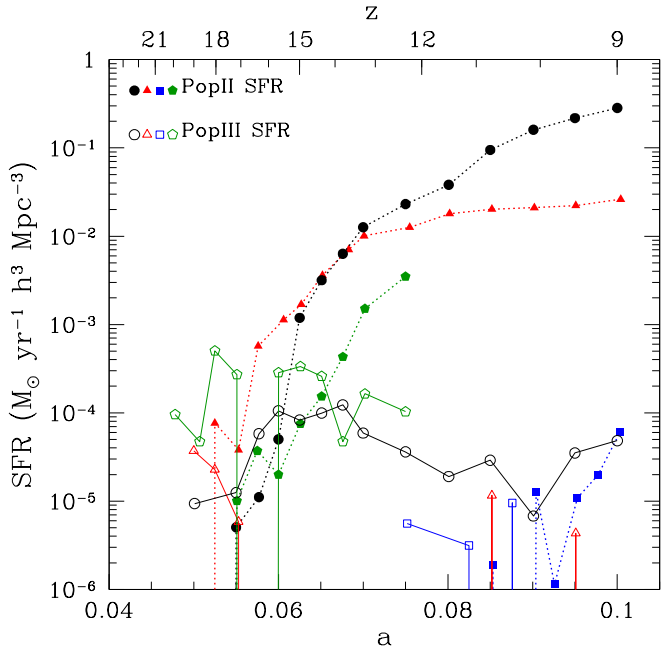}}
\vspace{-0.2cm}
\caption{ The SFR density vs scale factor for runs UnderDense$-$\_nH1e4\_fid (blue squares), UnderDense+\_nH1e4\_fid (red triangles), OverDense\_nH1e4\_fid (black circles), and OverDense\_HMpc\_HiRes (green pentagons). Global SFRs are significantly affected by the initial conditions in their respective simulation boxes. Pop III SFRs are denoted by open circles, Pop II SFRs by filled circles. }
\vspace{0.3cm}
  \label{fig:SFR1}
\end{figure}

Figure \ref{fig:MF} shows the number of halos capable of hosting Pop III stars. Box OverDense clearly dominates over the other two across the entire mass range, and since halo mass strongly correlates with the density of central gas cells, Box OverDense is host to many more star forming galaxies. 
Despite having fairly similar DC mode values, Box UnderDense$-$ and Box UnderDense+ have disparate halo abundances at the early epochs considered in our study. This is particularly visible at the high-mass end, where the most massive halo in Box UnderDense$-$ has about 10 analogs in Box UnderDense+. The H Mpc box contains fewer halos than Box UnderDense+, but per unit volume it contains higher density of massive halos, consistent with its higher $\Delta_{DC}$ value. At the time of formation of the first stars (around $a=0.05$), none of the halos are more massive than $2 \times 10^7 \, \Msun$. Figure \ref{fig:SFR1} presents the star formation rate (SFR) density in each box for the runs with fiducial parameters. We can immediately see that SFRs vary by orders of magnitude among the three boxes.
Only two galaxies are able to form Pop III stars in the extreme void represented by Box UnderDense$-$, and the total mass of Pop II is only $2300 \, \Msun$ by $a = 0.1$ ($z=9$) in this run. Such a narrow margin of error suggests that it is possible that with slightly different parameters for the initial overdensity, galaxies in this box may have never been able to form Pop III stars before all halos were stripped of gas by external ionizing radiation.

Box UnderDense+ and Box OverDense have similar global Pop II SFRs at $a < 0.07$ ($z>13$), as each box is being dominated by only a few galaxies at early times. However, the disparity between the mean density of the boxes becomes evident at late times, as Box OverDense becomes filled with a population of massive halos able to host dense cores in which Pop III, and subsequently, Pop II star formation is initiated. Box UnderDense+ continues hosting only one such galaxy for a large fraction of the duration of the simulation until $a=0.085$, when an external halo of primordial composition reaches the Pop III threshold density and later merges with the central galaxy. 

In contrast to the large differences between the simulation boxes, we find the variance in the SFR between various realizations within the same box to be low. Differences between the setup of our test runs in Paper I primarily affect Pop III stars, but at $a=0.055$, the total number of PopIII stars in Box OverDense was the same. Furthermore, Pop II stars constitute most of the star formation at $a > 0.07$. Once a given galaxy transitions to Pop II as the dominant stellar population, details of Pop III star formation do affect the history of that galaxy. The Pop III SFR stays relatively constant between $10^{-5}$ and $10^{-4} \, \Msunyr$ throughout the entire simulation. The range of values and relative constancy of this SFR up to $z$\textasciitilde$6$ are similar to the results found in larger-scale (4 Mpc) SPH simulations by \citet{johnson_etal13}. 

The H Mpc box, represented here by run OverDense\_HMpc\_HiRes, samples a very overdense region of the universe. The Pop III SFR density in this run significantly exceeds the other boxes before $a=0.065$, but the Pop II SFR density never reaches the corresponding values for Box OverDense and Box UnderDense+. This disparity results from the fact that the 1 Mpc boxes ultimately sample more massive galaxies which are able to sustain high Pop II star formation rates. 
Instead, a large fraction of low-mass galaxies simulated in the H Mpc box have considerable gas blowout after PISNe, delaying Pop II star formation for a cosmologically significant period of time.

\subsection{The Ejection and Gobbling of Pop III Metals}
The initial enrichment of the galactic ISM in our simulations happens almost exclusively through internal Pop III SNe rather than external intergalactic winds. This means that every galaxy will first have a phase during which Pop III stars constitute the entirety of the galactic stellar mass and drive all feedback. As our model considers only PISNe as a source of metal feedback during this Pop III phase, the enrichment required to transition to Pop II star formation is accomplished by at least one energetic PISN explosion. The explosion can be particularly potent, as it occurs in a hot, ionized, and diffuse medium carved out by the Pop III star's enhanced ionizing radiative feedback. The explosion therefore disrupts, or at least displaces, the dense gas necessary for further star formation.

\begin{figure}[t]
\vspace{-0.2cm}
\centerline{\epsfxsize3.5truein \epsffile{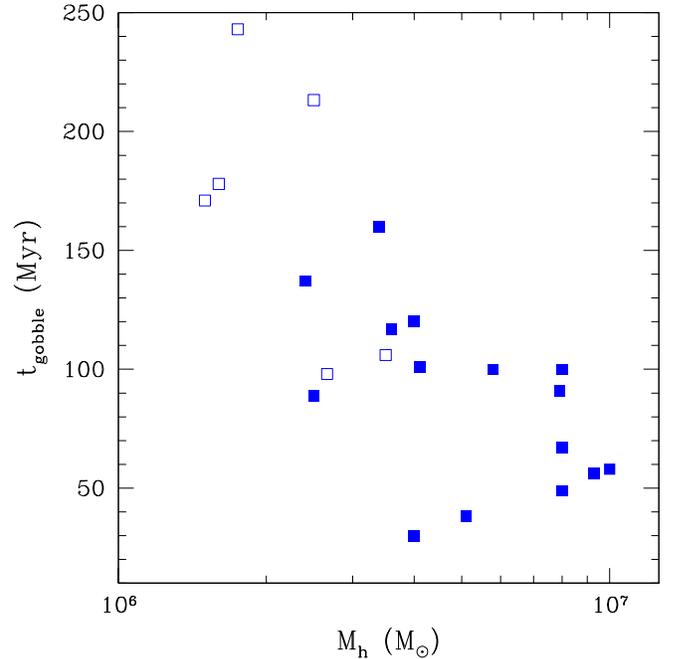}}
\vspace{-0.0cm}
\caption{ The timescale for a galaxy to re-accrete ("gobble up") 80\% of the metals ejected by its PISN vs. halo mass at the time of the explosion. Open squares represent instances where the primary galaxy merges with a secondary that has already been enriched by internal star formation, preventing accurate tracing of the metals initially associated with the primary galaxy's PISN. The more massive the galaxy, the shorter is the gobble timescale. Plotted galaxies are from runs OverDense\_nH1e4\_fid, OverDense\_SL7, and OverDense\_HMpc\_HiRes. }
\vspace{0.3cm}
\label{fig:gobble}
\end{figure}

We refer readers to Paper I, as well as prior work (e.g. \citealt{whalen_etal08}) for a detailed description of this process. Here, we note that the efficacy of the feedback of these individual massive stars depends significantly on the galaxy mass at the time of Pop III star formation. A \textasciitilde $10^6 \, \Msun$ "minihalo" can lose a significant fraction of its baryons, while a more massive (\textasciitilde $10^7 \, \Msun$) halo can better withstand the explosion. Most galaxies, particularly those in overdense environments, benefit from continued accretion from filaments, and can re-accrete the lost metals and baryons even if they were initially ejected beyond the virial radius. Particularly, many galaxies "gobble up" the metals that they previously ejected through a combination of rapid growth of their virial radius and true gravitational fallback of the ejecta. The virial radius of a given galaxy grows both because of continual buildup of matter in the outer parts of galaxies, and because the virial radius depends on the critical density of the universe, which evolves with redshift. In Figure \ref{fig:gobble}, we show the time it takes each galaxy to gobble up 80\% of the mass of metals generated in the initial PISN. We confirm that this timescale is relatively rapid (50-100 Myr) for galaxies more massive than the $3 \times 10^6 \, \Msun$ threshold we discovered in Paper I. In galaxies which are sufficiently massive (above \textasciitilde $10^7 \, \Msun$) at the time of the explosion, the metals rarely travel further than the virial radius, hence the gobble timescale is short. 

Galaxies below the $3\times 10^6 \, \Msun$ threshold have longer gobble timescales, and occasionally will merge with a more massive galaxy prior to the completion of the re-accretion process. Such instances prevent accurate tracing of the metals associated with the original PISN of the minihalo, and are therefore lower limits for the gobble timescales (denoted by open squares in Figure \ref{fig:gobble}). In particularly underdense environments with slow filamentary accretion, the metals are permanently ejected from the minihalo. In run OverDense\_HMpc\_HiRes, which resolves galaxies in the minihalo regime, only 20\% of PISNe result in such permanent ejections. The metals that are the result of the permanent ejections do not pollute other halos sufficiently to initiate Pop II star formation over the course of our simulations. Though these metals enrich the IGM, the total fraction of volume that is enriched beyond the critical metallicity, $\log_{10} Z/Z_\odot = -3.5$, has a peak value of 0.05\% at $z\approx10.5$ in our most minihalo-dominated simulation, run OverDense\_HMpc\_HiRes. 

The scatter of the gobble timescale for galaxies of a given mass is largely associated with their diverging mass accretion histories. We investigated the accretion history in both absolute and relative terms. The absolute growth rate is computed as the change in halo mass in the 100 Myr following the supernova. The relative growth is computed as the timescale for the halo to double the virial mass it had at the time of the explosion. We find that the absolute growth rate serves as a better predictor for the gobble timescale: all galaxies with ${dM}/{dt} > 0.2 \, \Msunyr$ have a gobble timescale of less than 100 Myr, while those with $0.04 < {dM}/{dt} < 0.2 \, \Msunyr$ show a wide range between 30 to 160 Myr. The 20\% of PISNe that result in permanent ejections have growth rates under $0.04 \, \Msunyr$ and mass doubling timescales longer than 90 Myr (most longer than 150 Myr).

\subsection{Transition to Normal Star Formation}

For the remainder of the paper, we mainly deal with galaxies that are more massive than $3\times 10^6\, \Msun$, and hence have relatively short gobble timescales. These galaxies typically dominate the star formation rate in the simulation boxes at late times. In particular, in this section, we focus on run OverDense\_nH1e4\_fid during the first 140 Myr after the first star forms. This corresponds to the range of scale factor $0.048<a<0.07$, or equivalently $20 > z > 13$. After that epoch the most massive galaxies in this box all form exclusively Pop II stars. Even more importantly, by $a = 0.07$, all Pop III stars in these galaxies have already exploded as SNe, and no longer provide radiative feedback. We will quantify the duration of the Pop III epoch in Section \ref{sec:EoE}.

\begin{figure}[t]
\vspace{-0.2cm}
\centerline{\epsfxsize3.5truein \epsffile{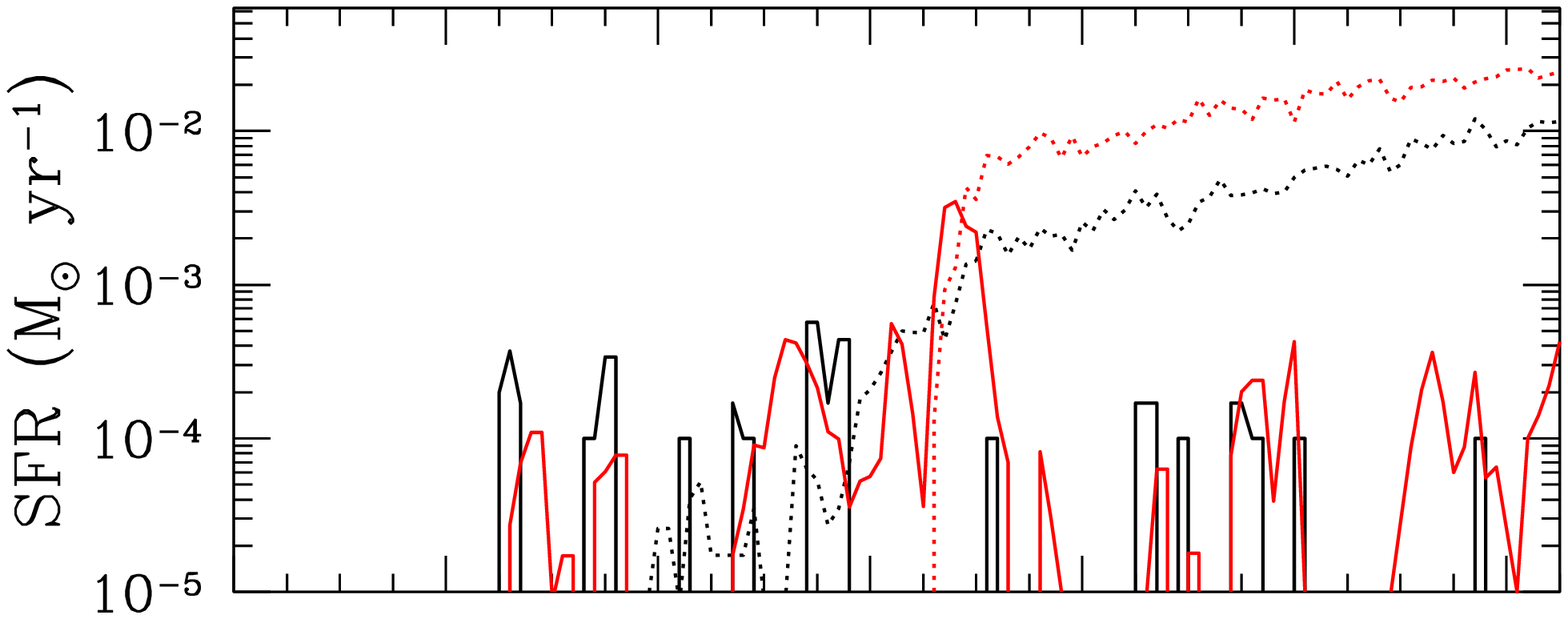}}
\vspace{-5.2cm}
\centerline{\epsfxsize3.5truein \epsffile{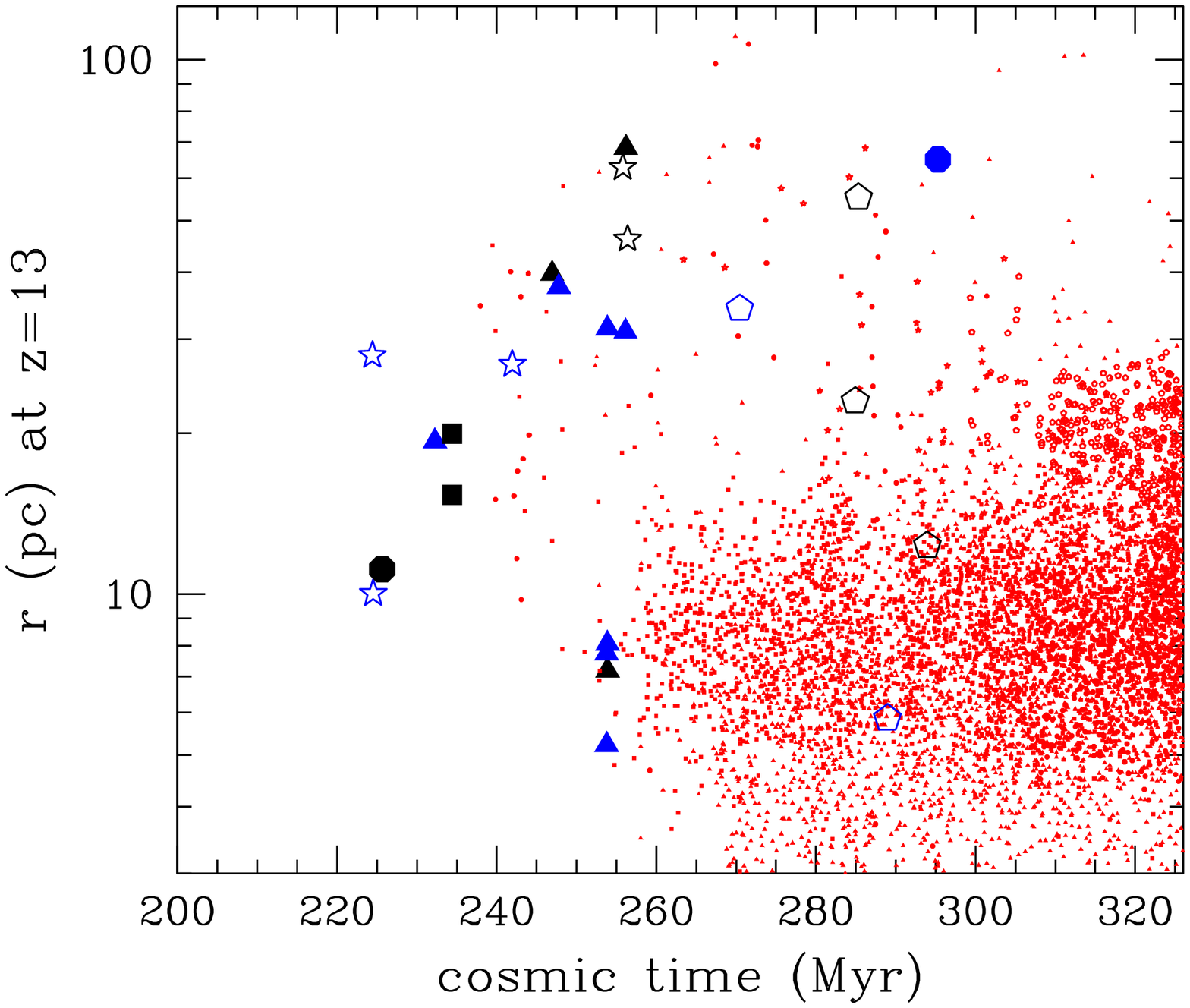}}
\vspace{-1.0cm}

\caption{ Bottom panel: distance of each star from its host galaxy's center at $a=0.07$ ($z=13$) vs the cosmic time when the star formed, for five of the most massive star-forming galaxies run OverDense\_nH1e4\_fid. Pop II stars are red points, the remnants of 100 $\, \Msun$ Pop III stars are blue symbols, and 170 $\, \Msun$ Pop III remnant tracer particles are black symbols. Individual galaxies are differentiated by different symbols. Every galaxy has a similar history: a few Pop III stars form and temporarily quench further star formation. However, eventually gas recollapses and self-sustained Pop II star formation begins within a 10-30 pc core, enriching the dense gas enough for Pop III star formation to be no longer possible within the galaxy. Pop III remnants often end up somewhat displaced from the Pop II core by as much as 100 pc. Top panel: the sum of SFRs for Pop III (solid) and Pop II (dotted) for these five galaxies, in black. For comparison, these rates are also plotted for run OverDense\_LowMass, in red. }
\vspace{0.3cm}
\label{fig:distage}
\end{figure}

Figure \ref{fig:distage} shows the distance of each stellar particle from the center of its host galaxy at $a = 0.07$ vs the cosmic time at which the stellar particle was formed. Five of the most massive star-forming galaxies are stacked together in this plot, showing the similarity in the assembly history of these galaxies. We see that each of the five galaxies does not host more than a few Pop III stars. If a $100 \, \Msun$ star is the first to form in a given galaxy, it will not release any metals, and further Pop III star formation will be possible after the gas recollapses. If on the other hand a $170 \, \Msun$ star is formed, Pop III star formation is rarely possible in the same galaxy again. The metals generated by a single PISN from a $170 \, \Msun$ star are enough to switch the galaxy's primary mode of star formation to be Pop II by pushing the metallicity of gas above the critical threshold $\log_{10} Z/Z_\odot = -3.5$ (as has been previously demonstrated e.g. \citealt{greif_etal10, wise_etal12a}). A few exceptions to this pattern happen when the sites of Pop III star formation within the galaxy are sufficiently spread apart (i.e. two Pop III stars form in two different halos that subsequently merge) or alternatively, the Pop III stars form sufficiently close together in time. The latter case is possible when many cells within the same giant molecular cloud reach the threshold density for star formation and several Pop III stars form before their feedback disperses the remaining gas in the cloud. This produces Pop III star multiples that are very close to each other in location and age.

The spatial distributions of Pop III stellar remnants and Pop II stars within their host galaxy are significantly different. Pop III stellar remnants can be found out to 70 pc from the center, while most Pop II stellar particles at this epoch are within 20 pc. Note that the centers of the galaxies are defined to only \textasciitilde5-15 pc, because of the limited number of DM particles and force resolution. High-mass galaxies have better-defined centers than low-mass galaxies. Both stellar populations generally form close to the center of the host galaxy, but Pop III stars typically form at earlier times when the galaxy is not very massive, and the spatial concentration of DM particles is not very high. In this environment, dense star-forming gas has a clumpy structure, which is off-center from the DM cusp. After powerful Pop III feedback ejects gas from this region, the Pop III remnants become less bound, further increasing the apocenter of their orbit. On the other hand, Pop II stars are rapidly produced only once the galaxy becomes relatively massive and concentrated, and the DM center is more closely associated with the location of the densest gas. Pop II thermal feedback and radiative feedback is not strong enough to shut off further Pop II star formation near the galactic center, ultimately leading Pop II stars to have a very high spatial concentration.

\subsection{Epoch of Equivalence}
\label{sec:EoE}
The duration of the Pop III epoch has been previously studied (e.g. \citealt{trenti_stiavelli09}) by comparing the SFRs of the two populations, and finding the epoch where the Pop III SFR drops off dramatically compared to Pop II. Our Figure \ref{fig:SFR1} shows that by $a=0.07$ ($z=13$), Pop II star formation is more common in all boxes except the void represented in Box UnderDense$-$. However this approach may not properly account for the impact Pop III stars have on the universe at early times, as Pop III stars are able to generate significantly more thermal, ionizing, and metal feedback per baryon compared to their Pop II counterparts (e.g. \citealt{tumlinson_shull00}). To more precisely quantify the impact of Pop III stars, we compute the relative "budget" of thermal energy, metals, and ionizing photons contributed in each galaxy by the two stellar populations, integrated over the lifetimes of all stellar particles. For the purpose of this analysis, we define the "epoch of equivalence" as the time when Pop II stars have generated just as much ionizing radiation as their Pop III counterparts. In turn, we define the "duration of the Pop III phase" as the length of time from the formation of the first Pop III star to the epoch of equivalence.

\begin{figure}[t]
\vspace{-0.2cm}
\centerline{\epsfxsize3.5truein \epsffile{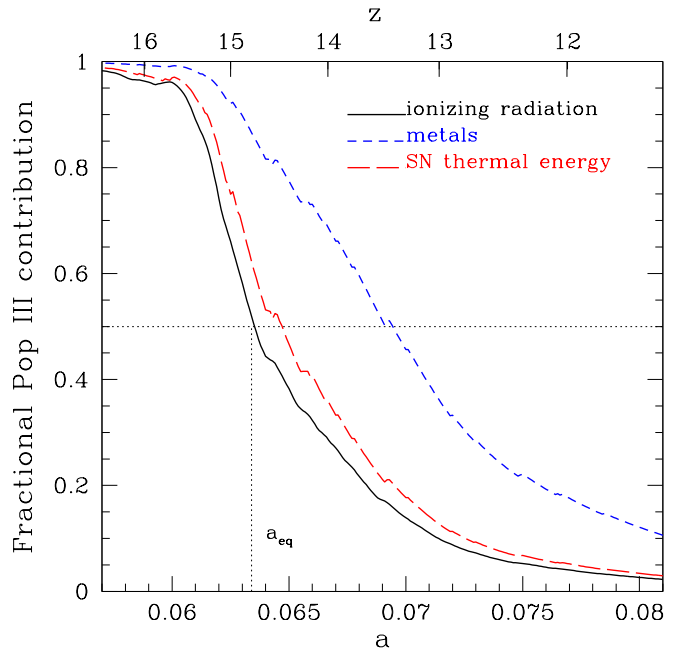}}
\vspace{-0.0cm}
\caption{ The fractional contribution of Pop III stars to the integrated feedback budget of the universe in run OverDense\_nH1e4\_fid vs scale factor. This fraction is calculated separately for metals (blue), ionizing radiation (black), and thermal radiation due to SNe (red).}
\vspace{0.3cm}
\label{fig:finalduration}
\end{figure}

A global picture of the feedback budget is shown in Figure \ref{fig:finalduration}. Here, we calculate at each epoch the relative contribution of Pop III stars to the budget of ionizing photons, thermal energy from SNe, and metals injected into the ISM in run OverDense\_nH1e4\_fid. Every budget can be examined individually. Pop II stars have produced as many ionizing photons as Pop III stars by $a=0.0633$ ($z=14.8$). The total thermal energy released by SNe in Pop II stars surpasses that produced by Pop III stars at $a=0.0642$ ($z=14.6$). The total mass of metals produced in Pop II stars surpasses that of Pop III at $a=0.0689$ ($z=13.5$). Though we have chosen to define the epoch of equivalence in terms of ionizing radiation, any one of these feedback quantities leads us to the conclusion that the duration of the Pop III phase is short. 

The exact timing of the epoch of equivalence can be affected by the simulation's ability to resolve low-mass halos that are susceptible to the deleterious effects of Pop III feedback, and hence have longer delays for the onset of Pop II star formation. The resolution necessary to probe such low-mass galaxies is not attained in run OverDense\_nH1e4\_fid. Run OverDense\_SL7 probes the same volume as run OverDense\_nH1e4\_fid, but Pop III stars form at earlier times, when the host halos are less massive. In this run, the epochs of equivalence shifts to $a=0.0633$, $a=0.0674$ and, $a=0.0710$ for ionizing radiation, SNe thermal energy, and metals, respectively. For run OverDense\_HMpc\_Hires, which resolves many galaxies in the minihalo regime, the epoch of equivalence for the three types of feedback is reached at $a=0.0721$, $a=0.0735$, and $a=0.0764$. These consistently small shifts in the duration of Pop III dominance give us no reason to believe that significant contributions of feedback from Pop III stars are missed in simulations that fail to resolve minihalos. 

We did not include Box UnderDense+ and UnderDense$-$ in this analysis, as they sample few star-forming galaxies and do not make significant contributions to the budget of either Pop III or Pop II feedback.

To understand the transition to normal star formation in individual galaxies, we also budget the feedback from stars within a given galaxy. All numbers quoted in this section reflect the integrated total feedback contributions over the lifetime of stellar particles up to the epoch considered. As a case study, we take the third galaxy to form a Pop III star in run OverDense\_nH1e4\_fid. The first star forms at $a=0.055$ ($z=17.2$), when the halo mass and gas mass of this galaxy are $9.8 \times 10^6 \, \Msun$ and $1.2 \times 10^6 \, \Msun$, respectively. At $a = 0.06$ ($z=15.7$, 23 Myr later) this galaxy has formed a total of $118 \, \Msun$ of Pop II stars. The first and only Pop III star that formed in the galaxy was a $170 \, \Msun$ PISN progenitor. Although the total mass of each stellar population is comparable, the one Pop III star still has had dominant influence on the galaxy by way of its stronger feedback. The Pop III star contributed 98\% of the metals, 88\% of the ionizing photons, and 93\% of the thermal energy. This galaxy is clearly in the phase of its evolution where the effects of Pop III star formation are most likely to be seen. However, we again note that despite this, the galaxy has not been significantly perturbed by the Pop III star's injection of energy. Only 1.7\% of its hydrogen is ionized, and all of the metals are confined in the gas well within the virial radius. 

By $a = 0.0675$ (80 Myr after the first star formed), one more Pop III star has formed, giving $270 \, \Msun$ of Pop III mass formed in the galaxy, but now the total Pop II mass has increased to $1260 \, \Msun$. This second Pop III star has a metallicity of $\log_{10} Z/Z_\odot = -4.7$, suggesting it formed in gas that was significantly enriched, but below the critical metallicity of $\log_{10} Z/Z_\odot = -3.5$. Therefore, whether the star should be Pop III or Pop II is rather sensitive to the particular setup of the model and stochastic effects. These two Pop III stars still contribute 91\% of the metals, 61\% of the ionizing photons and 71\% of the thermal energy. We see that the energy contributions from Pop II stars are catching up with those of Pop III stars, but the metal contribution of Pop III stars remains dominant. This confirms the trend of Figure \ref{fig:finalduration} that the large yield of metals produced by PISNe is the longest lasting contribution of Pop III star formation. 

By $a = 0.075$ ($z=12.3$, 134 Myr after the first star formed), the epoch of Pop III stars is clearly over in this galaxy despite the fact it has merged with another galaxy, which also hosted a recent PISN progenitor Pop III star. The combined initial mass of Pop III stars is now $440 \, \Msun$, while Pop II stars contribute $48000 \, \Msun$. The Pop III stars still contribute 26\% of the metal budget, but a mere 5\% of the ionizing photon budget and 8\% of the thermal energy budget. By $a = 0.08$, the Pop III metal contribution has finally dwindled away to only 8\%.

The epoch when Pop III stars have affected the evolution of the galaxy is thereby confined to between the time when the first star formed (around $a \approx 0.055$) and the time when the budget of energy and metals of Pop II stars begins to dominate over Pop III stars ($a \approx 0.07$). The window of Pop III dominance only corresponds to $\Delta a \approx 0.015$, or 100 Myr.
 
 \begin{figure}[t]
\vspace{-0.2cm}
\centerline{\epsfxsize3.5truein \epsffile{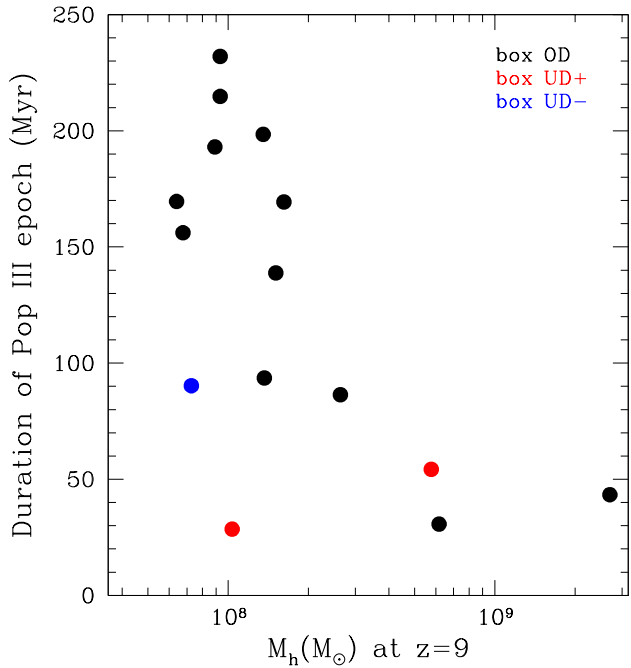}}
\vspace{-0.0cm}
\caption{ The duration of the epoch of Pop III dominance vs halo mass of host galaxies at $a = 0.1$ ($z=9$). Plotted are galaxies in run OverDense\_nH1e4\_fid (black), run UnderDense+\_nH1e4\_fid (red), and run UnderDense$-$\_nH1e4\_fid (blue). More massive galaxies have shorter Pop III epochs.}
\vspace{0.3cm}
\label{fig:Mh_Duration}
\end{figure}

Figure \ref {fig:Mh_Duration} shows the duration of Pop III stars in this galaxy as well as in others, plotted against halo mass at $a=0.1$ ($z=9$). We note that the length of the Pop III epoch is generally between 20-200 Myr, confirming that the galaxy discussed in the above paragraphs is not unique. The figure also reveals a trend where more massive galaxies exit the Pop III stage earlier. These galaxies form in more biased regions, where the infall of fresh gas overcomes the negative feedback generated by the thermal and ionizing radiation of Pop III stars, and Pop II star formation erupts in the newly metal-enriched gas. Galaxies with higher mass at the time of Pop III star formation can also radiate the PISN energy more efficiently. 

\begin{figure}[t]
\vspace{-0.2cm}
\centerline{\epsfxsize3.5truein \epsffile{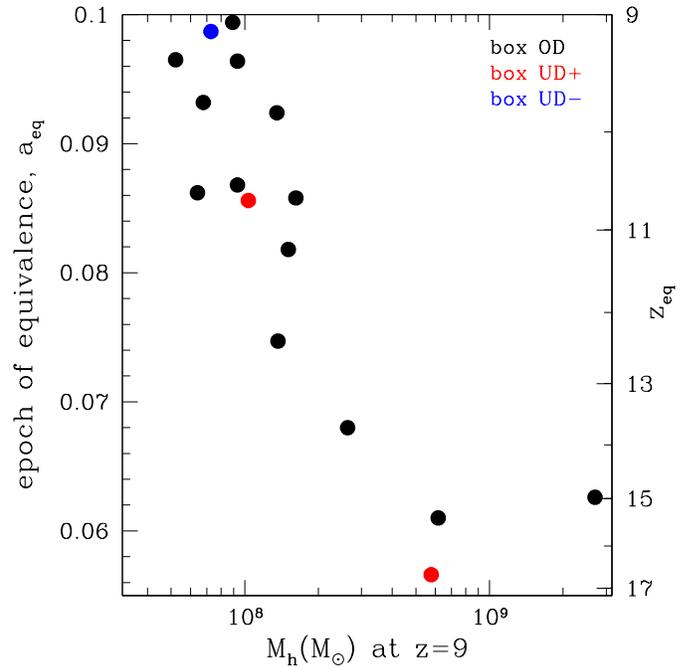}}
\vspace{-0.0cm}
\caption{ The epoch at which the contributions of Pop III and Pop II stars to the cumulative ionizing photon budget are equal vs. halo mass of host galaxies at $a = 0.1$ ($z=9$). Plotted are galaxies in run OverDense\_nH1e4\_fid (black), run UnderDense+\_nH1e4\_fid (red), and run UnderDense$-$\_nH1e4\_fid (blue). The epoch of equivalence is earlier in more massive galaxies.}
\vspace{0.3cm}
\label{fig:Mh_aequiv}
\end{figure}

Figure \ref{fig:Mh_aequiv} shows that the epoch of equivalence for individual galaxies is also strongly correlated with the mass of the halo at $a = 0.1$ ($z=9$). This trend is perhaps not surprising, as the sites of formation for the first stars have to be the most biased regions of the simulation box, which would also produce the most massive galaxies. It is impressive, however, that the trend is so definitive that it suggests we can predict the epoch at which Pop III stars became subdued based simply on the halo mass.

We have also calculated the duration of the Pop III epoch by using the metal and thermal energy budgets instead of the ionizing photon budget. The thermal energy dominance of Pop III stars tends to last on average about \textasciitilde10 Myr longer than the ionizing photon epoch, with little variance, while Pop III metals remain dominant for \textasciitilde20-50 Myr longer. These differences are significant, and can be understood by looking at the relative efficiency of metal production of $170 \, \Msun$ Pop III stars compared to Pop II counterparts. While our Pop II stellar particles only generate an amount of metals equivalent to 1.1\% of their initial mass, $170 \, \Msun$ Pop III stars release almost 47\% of their mass in metals following a PISNe, meaning that Pop III metal feedback is \textasciitilde40 times more effective. In comparison, the number of ionizing photons per stellar baryon per lifetime is 6,600 for Pop II stars vs. 34,500 for $170 \, \Msun$ Pop III stars respectively, or only a factor of \textasciitilde5 difference in efficiency. The variation between individual galaxies can be caused by the number of non-enriching $100 \, \Msun$ Pop III stars that form there, and by the rate at which Pop II stars form, as their feedback is released gradually over a longer interval of time. Despite this variation associated with using different feedback budget quantities, our conclusion on the brevity of the Pop III epoch within each galaxy remains robust. 

\subsection{Extreme Pop III Feedback}

\label{sec:extremeFB}
The duration of the phase of Pop III dominance may depend on the details of the implementation of stellar feedback. Here, we consider two additional runs with artificially enhanced Pop III feedback. By $a=0.0625$ ($z=15$), the fiducial run OverDense\_nH1e4\_fid has 34 Pop III stars, while run OverDense\_ExtremeSN has 28 stars and the OverDense\_ExtremeRad run has 19 stars. At least part of this variation can be explained by the stochastic effect that the OverDense\_ExtremeRad run happened to form $170 \, \Msun$ PISNe progenitors more frequently than the other runs, hence quenching further Pop III star formation in their hosts. However, other evidence suggests that the extreme radiation does indeed quench star formation very efficiently. This can be discerned by examining Pop II star formation, as $20000 \, \Msun$ formed in run OverDense\_nH1e4\_fid, $7000 \, \Msun$ formed in run OverDense\_ExtremeSN, and $4000 \, \Msun$ in run OverDense\_ExtremeRad at $a=0.0625$. While we showed in Paper I that the effects of baryon expulsion in individual galaxies are temporary in both extreme feedback runs as well as the fiducial run, the cumulative effect of many more extreme feedback events appears to have suppressed of star formation at early times. More specifically, the relatively high IGM temperature in run OverDense\_ExtremeRad (discussed in Paper I) means that newly accreted gas in this run will take longer to cool and condense to the densities required for star formation. In our models, radiative feedback is more effective than SN thermal feedback as a means of regulating star formation and modifying the structure of the ISM and IGM. 

In both extreme feedback runs, the epoch of equivalence is pushed back relative to the fiducial run, however it is always reached before $z=13$. The epoch of equivalence for ionizing radiation increased by $\delta a = 0.0073$ in run OverDense\_ExtremeRad, while the epoch of equivalence for SN thermal energy increased by $\delta a = 0.0068$ in run OverDense\_ExtremeSN. Even the drastic modifications to Pop III feedback output causes only marginal shifts in the duration of their epoch of dominance.

\subsection{Low Mass Pop III IMF}

\label{sec:LowMass}

Using a standard \citet{miller_scalo79} IMF for Pop III stars causes the duration of the Pop III epoch to be somewhat shorter than in the top-heavy fiducial case. The epoch of equivalence for ionizing radiation is reached in run OverDense\_LowMass at $a=0.0622$, a mere 90 Myr after the first star forms. 

However, the more important distinction in run OverDense\_LowMass is the relative lack of suppression of star formation within individual galaxies. By $a=0.065$ ($z=14.4$), the total stellar mass formed in run OverDense\_LowMass is a factor of \textasciitilde 3 larger than in the fiducial run OverDense\_nH1e4\_fid (see top panel of Figure \ref{fig:distage}). The majority of stellar mass in both runs is contained in the three most massive star-forming galaxies, each of which has more stellar mass in the run OverDense\_LowMass. In this run, nearly all of the stellar particles that formed in these three galaxies are within 10 pc of their galaxy's center, enriching the dense gas slowly and steadily to perpetuate further star formation. A similar Pop II core is seen in the fiducial run, however some Pop II stars and Pop III stellar remnants are found up to 100 pc from their galactic centers (see bottle panel of Figure \ref{fig:distage}).

\section{Discussion and Conclusions}

We find that Pop III star formation can be expected to extend until at least $z \approx 10$. However, Pop III stars are not the dominant source of any form of global feedback past $a\approx0.07$ ($z\approx13$) in our models, and only dominate individual galaxies for 20 to 200 Myr. 

The UV radiation of Pop III stars and their powerful supernovae can temporarily evacuate neutral gas from the first galaxies, creating an opportunity for ionizing flux from the subsequent generations of Pop II stars to escape into the IGM. However, it is unlikely that the escape fraction was high because, as we showed in Paper I, accretion from filaments brings back most of the expelled gas on roughly the same timescale (\textasciitilde150 Myr) as the duration of the Pop III phase. This means that by the time star formation resumes at the center, the entire galaxy will again be surrounded by neutral gas, making it difficult for the ionizing photons to escape.

We also showed in Paper I that most of the metals generated in PISN would eventually fall back into the host galaxy rather than remaining in the IGM. This fact, combined with the relative insignificance of the Pop III metal budget, leads us to conclude that Pop III stars did not play much of a role in enriching the IGM, and that instead enrichment happened by normal stellar populations during and after the epoch of reionization. Though Pop III star formation is self-limiting in individual galaxies, the absence of universal enrichment implies that Pop III stars could form in underdense regions long after the universe has primarily transitioned to normal star formation. It is in the least massive galaxies that formed in underdense regions where we might expect to find the most recent signatures of Pop III stars, suggesting that ultra-faint dwarf spheroidal galaxies and their disrupted remnants within the Milky Way stellar halo may be the best place to look. As the chemical signature of PISNe metals is expected to be different from that of Type II SNe, and the metal budget of Pop III stars is the last feedback tracer to be overtaken by Pop II, studying the spectra of old metal-poor stars with peculiar chemical abundances may be one of the most promising ways to observationally constrain the chemical history of the first galaxies. The first efforts in this field of 'stellar archeology' have already yielded interesting results (e.g. \citealt{frebel_etal07, frebel_bromm12}), but there is more work to be done to increase the sample of metal-poor stars, and to improve our knowledge of their abundances, many of which are currently too faint for detailed spectroscopy. The next generation of spectroscopic and photometric surveys, such as Gaia-ESO and SkyMapper, will significantly improve the sample, while the light collecting area of upcoming large ground-based optical telescopes will enable detailed follow-ups of significantly fainter candidates \citep{frebel11}.

According to mock observational analysis for JWST by \citet{pawlik_etal12}, galaxies with SFR in excess of $0.1 \Msunyr$ may be detectable at $z>10$. Only two galaxies in our simulation box meet this criterion in the final snapshot at $z=9$, and both are evolved far beyond the point where Pop III stars made a significant contribution to their feedback budgets. According to a separate analysis by \citet{zackrisson_etal12}, JWST surveys that target gravitationally lensed fields are much more likely to discover primordial galaxies than ultra-deep unlensed fields. These authors suggest that galaxies in which 0.1\% of baryons are in the form of Pop III stars are detectable through direct starlight if they are in the lensed field of the galaxy cluster J0717.5+3745. However, none of our galaxies meet this criterion at any given time.

A more promising way to detect galaxies that are still dominated by Pop III stars is by looking for PISN events directly. The light curve of a PISN is thought to have a similar peak luminosity to SN Ia, but can remain bright for hundreds of days, producing a distinct signature (e.g. \citealt{scannapieco_etal05}). This idea has been explored and shown to be feasible for both LSST \citep{trenti_etal09} and JWST \citep{hummel_etal12, pan_etal12a, whalen_etal12}. Again, considering our findings that underdense regions of the universe may remain pristine until late times, we conclude that the best approach for observing campaigns is to focus on areas where the spatial clustering signal is low, such as cosmic filaments that connect two bright galaxies, to maximize the number of bright, low-redshift detections. 

If Pop III stars instead formed with a normal IMF (e.g. \citealt{greif_etal11}), they would have even less impact on the universe, as demonstrated by our low-mass IMF run. On the other hand, while our extreme feedback runs showed that the Pop II transition could be delayed with stronger Pop III feedback parameters, the transition inevitably happened before $z=13$ and takes less than 250 Myr for even the least massive galaxies. In both cases, our modifications to the feedback output were simplistic ways to test the range of potential impact of Pop III stars without detailed modeling of additional physical processes. However, it is becoming clear that missing physics in galaxy formation simulations can have a wide variety of effects that are not accounted by our limited approach. For example, momentum-driven winds from radiation pressure can significantly stir the ISM increasing its susceptibility to other forms of feedback \citep{wise_etal12b, agertz_etal13}. The presence or absence of dust can affect the shielding capabilities of molecular clouds from UV radiation, thereby modifying fragmentation and potentially the IMF of stars \citep{aykutalp_spaans11}. Dark matter annihilation \citep{smith_etal12}, cosmic rays \citep{jasche_etal07,stacy_bromm07, uhlig_etal12}, magnetic fields \citep{turk_etal12}, and X-ray feedback from the first black holes \citep{haiman11, jeon_etal12} could also be potentially important. We plan to explore them further in the next generation of simulations. 

\acknowledgements 
We thank the anonymous referee for useful comments which helped to improve the quality of this work. A.L.M. acknowledges the support of the Rackham pre-Doctoral Fellowship awarded by The University of Michigan. O.Y.G. was supported in part by NSF grant AST-0708087 and NASA grant NNX12AG44G. This work was supported in part by the DOE at Fermilab and by the NSF grant AST-0708154. M.Z. is supported in part by a 985 grant from Peking University.

\makeatletter\@chicagotrue\makeatother

\bibliography{FS}

\end{document}